# Surface-specific vibrational spectroscopy of interfacial water reveals large pH change near graphene electrode at low current densities

*Yongkang Wang,[1,2,#] Takakazu Seki,[2,#] Xuan Liu,[3] Xiaoqing Yu,[2] Chun-Chieh Yu,[2] Katrin F. Domke,[2,4] Johannes Hunger,[2] Marc T. M. Koper,[3] Yunfei Chen,[1,\*] Yuki Nagata,[2,\*] and Mischa Bonn,[2,\*]*

[1] School of Mechanical Engineering, Southeast University, 211189 Nanjing, China.
[2] Max Planck Institute for Polymer Research, Ackermannweg 10, 55128 Mainz, Germany.
[3] Leiden Institute of Chemistry, Leiden University, Einsteinweg 55, 2333CC Leiden, The Netherlands.
[4] University Duisburg-Essen, Faculty of Chemistry, Universitätsstraße 5, 45141 Essen, Germany.
[#] Yongkang Wang and Takakazu Seki contributed equally to this work.

*Correspondence to: yunfeichen@seu.edu.cn, nagata@mpip-mainz.mpg.de, bonn@mpip-mainz.mpg.de

## Abstract

Molecular-level insight into interfacial water at buried electrode interfaces is essential in elucidating many phenomena of electrochemistry, but spectroscopic probing of the buried interfaces remains challenging. Here, using surface-specific vibrational spectroscopy, we probe and identify the interfacial water orientation and interfacial electric field at the calcium fluoride ($CaF_2$)-supported electrified graphene/water interface under applied potentials. Our data shows that the water orientation changes drastically at negative potentials (<-0.03 V *vs*. Pd/$H_2$), from O-H group pointing *down* towards bulk solution to pointing *up* away from the bulk solution, which arises from charging/discharging not of the graphene but of the $CaF_2$ substrate. The potential-dependent spectra are nearly identical to the pH-dependent spectra, evidencing that the applied potentials change the local pH (more than five pH units) near the graphene electrode even at a current density below 1 $\mu A/cm^2$. Our work provides molecular-level insights into the dissociation and reorganization of interfacial water on an electrode/electrolyte interface.

## Introduction

Water at a potentiostatically controlled electrode surface is relevant for a wide range of scientific and technological systems.[1–5] Homogeneous dielectric continuum models assumed in traditional mean-field theories often fail at the electrode/electrolyte solution interface[1,6–8] in which arrangements of interfacial water molecules are heterogeneous,[9] the water response to the electric field is supposed to be asymmetric,[10,11] and dissociation of water molecules, also not contained in continuum models, plays an important role.[2,12] For instance, the reorganization of interfacial water molecules and their spatial arrangement are closely linked with capacitive charge storage,[3,13] and with the electron transfer mechanism across electrode/water interfaces, such as in electrocatalytic water splitting.[2,12,14] Furthermore, changes in the local pH near the electrode with the applied potential affect the chemical reactions [15–17] because the changing local pH can significantly perturb the interfacial electric field,[2,18] and, in turn, the arrangement of interfacial water. As such, elucidating these local molecular events is essential for understanding the electrochemistry at electrode/water interfaces.

To probe the interfacial water structure, heterodyne-detected sum-frequency generation (HD-SFG) spectroscopy is ideal, owing to three advantages: interface specificity, molecular specificity, and orientation specificity.[19,20] The SFG signal can only originate from the interface owing to its selection rule that centro-symmetry must be broken for non-zero SFG signals. As such, it naturally excludes contributions from the bulk, allowing us to probe the interfacial response selectively.[19,21] An SFG signal is enhanced when the infrared (IR) frequency is resonant with the molecular vibration, providing molecular specificity. Furthermore, HD-SFG signals provide (i) the absolute orientation of molecules (*up-/down*-orientation),[22,23] and (ii) the surface charge, and thus the interfacial electric field.[24,25] However, applying HD-SFG spectroscopy to buried electrode/water interface is challenging because IR often cannot reach the buried interfaces.[26–28] This is enabled by using graphene as an electrode; graphene, an atom-thickness metal-like material, can serve as the electrode while allowing the IR light to reach the electrode surface.[11,29–32] As such, combining SFG with a graphene electrode allows us to explore the molecular-level insight into the interfacial conformations under electrified conditions.

Here, we perform HD-SFG measurements at the $CaF_2$-supported electrified graphene/water interface. We observe that the $\text{Im}\left(\chi_{\text{eff}}^{(2)}\right)$ spectra of the O-H stretching mode of water drastically change in between -0.63 V and -0.03 V versus the $Pd/H_2$ electrode. We find that it is not due to the variation of charges on the graphene, but due to the drastic surface charge variation of the $CaF_2$ substrate, likely accommodated by water trapped in between the $CaF_2$ and the graphene. From HD-SFG measurements at various pH conditions, we identify that the surface charge of the $CaF_2$ substrate varies substantially even at 1 $\mu A/cm^2$ current density due to the HER-induced local pH change at the $CaF_2$/graphene interface. This work provides molecular-level insights into the dissociation and reorganization of interfacial water on a potentiostatically controlled electrode surface.

## Experimental

***Sample preparation.*** The schematic diagram of the experimental setup for the *in-situ* HD-SFG measurement under controlled electrochemical conditions is shown in Fig. 1a. A large-area CVD-grown monolayer graphene was transferred onto a $CaF_2$ window and was used as the working electrode (WE), similar to previous studies.[11,32] The prepared sample was then mounted on a homemade electrochemical flowing cell, which enables *in-situ* HD-SFG measurement under controlled electrochemical conditions. As the counter electrode (CE) and reference electrode (RE), gold, and $Pd/H_2$ wires were used, respectively. We used 1 mM $NaClO_4$ aqueous solution for Raman measurements, while we used 1 mM and 10 mM solutions for electrochemical measurements, and 1 mM and 100 mM solutions for the SFG measurements. The solution pH was controlled by adding HCl or NaOH to the solution. More details about the graphene transfer and characterization, electrode preparation, and electrochemical cell design can be found in the *Supporting Information (SI)*.

***HD-SFG setup.*** HD-SFG measurements were performed using a non-collinear beam geometry with a Ti:Sapphire regenerative amplifier laser system (Spitfire Ace, Spectra-Physics, centered at 800 nm, ~40 fs pulse duration, 5 mJ pulse energy, 1 kHz repetition rate). A part of the output was directed to a grating-cylindrical lens pulse shaper to produce a narrowband visible pulse, while the other part was used to generate a broadband IR pulse through an optical parametric amplifier (Light Conversion TOPAS-C) with a silver gallium disulfide ($AgGaS_2$) crystal. We focused the IR and visible beams non-collinearly onto a ZnO film deposited on a $CaF_2$ window to generate a local oscillator (LO) signal, similar to the reference.[33] The LO signal passed through a phase modulator. After that, IR, visible, and LO beams were re-focused onto the graphene/water interface at angles of incidence of 33°, 39°, and 37.6°, respectively. The measurements were performed under a nitrogen atmosphere at *ssp* polarization combination, where *ssp* stands for *s*-polarized SFG, *s*-polarized visible, and *p*-polarized IR beams. More details about the HD-SFG measurement can be found in the *Supporting Information (SI)*.

## Results and discussion

To understand the structure of the interfacial water on the electrified graphene surface, we measured $\mathrm{Im}(\chi_{\mathrm{eff}}^{(2)})$ spectra in the O-H stretching mode frequency region (2900-3700 cm$^{-1}$). The data at different potentials with respect to the reversible hydrogen electrode (RHE, Pd/H$_2$) are displayed in Fig. 1b. At +0.57 V, the $\mathrm{Im}(\chi_{\mathrm{eff}}^{(2)})$ spectrum exhibits a negative band spanning from 2950 cm$^{-1}$ to 3550 cm$^{-1}$. This negative band is assigned to the O-H stretching mode of water molecules hydrogen-bonded (H-bonded) to the other water molecules,[19,34] and the negative sign of this band indicates that the H-bonded O-H group points *down* (towards the bulk solution).[23,35] This negative band contribution is insensitive to the variation of the applied potential in the range of +0.57 V to -0.03 V.

When we decreased the potential from -0.03 V to -0.63 V, the sign of the H-bonded O-H stretching band flips from negative to positive, illustrating that the orientation of the interfacial water molecules changes from *down* to *up* (H-bonded O-H pointing away from the bulk solution). During this transition, we further observed the appearance of a negative 3630 cm$^{-1}$ O-H stretch peak. This observation is surprising because previously, the sign of this peak was inferred as positive from homodyne-detected SFG measurement and was assigned to the O-H group of water interacting with the CaF$_2$-supported graphene.[11] If this is the case, this O-H group points *up* to the CaF$_2$-supported graphene sheet, providing a positive 3630 cm$^{-1}$ peak.[35] The sign of the 3630 cm$^{-1}$ peak is, however, negative, indicating that the O-H group rather points *down* towards the bulk solution. This mismatch shows that the peak does not arise from the O-H group of the topmost interfacial water molecules. Instead, the negative 3630 cm$^{-1}$ peak can be assigned to the O-H stretch of the Ca-O-H species on the CaF$_2$ surface,[36] because the O-H group of the Ca-O-H species points *down* towards the bulk solution. From -0.43 V to -0.63 V, both the positive H-bonded O-H stretch and the negative Ca-O-H stretch peaks were further enhanced. From -0.63 V down to -1.23 V, the $\mathrm{Im}(\chi_{\mathrm{eff}}^{(2)})$ spectra were found to be again insensitive to the applied potentials.

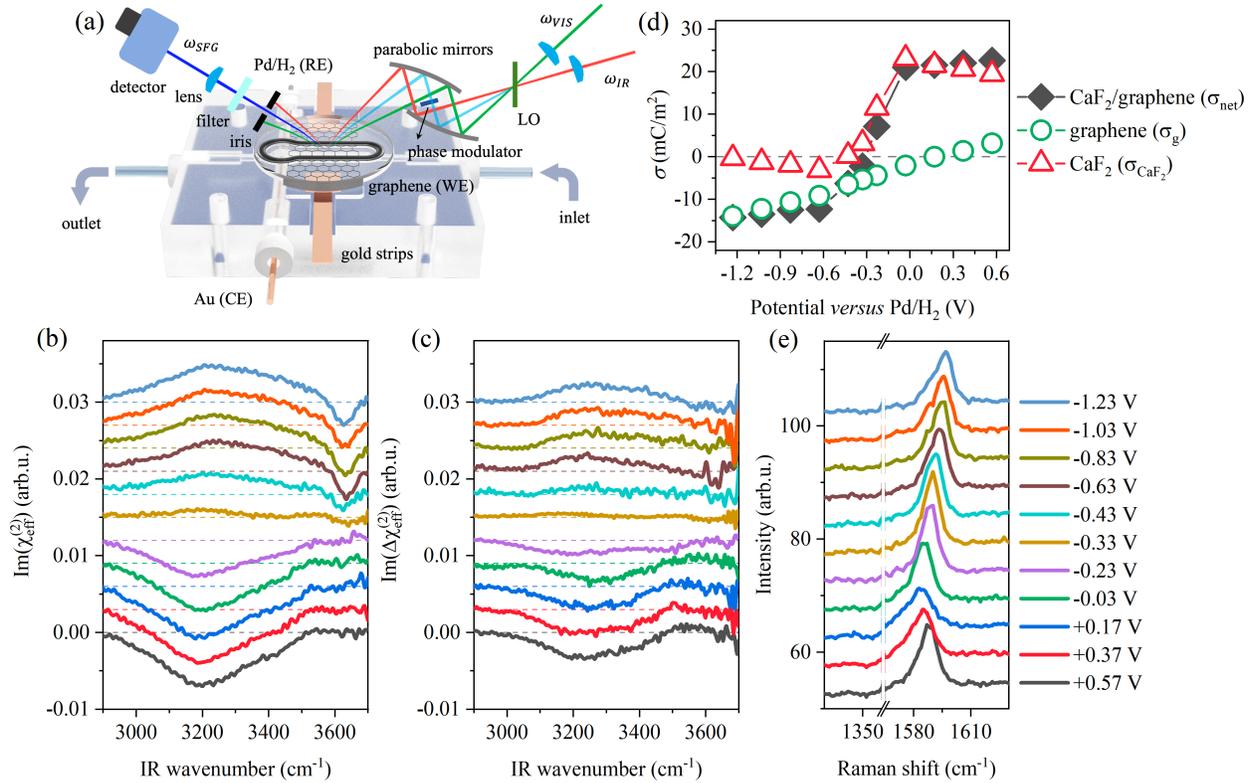

**Figure 1. O-H stretching spectra at the CaF$_2$-supported graphene/water interface measured by HD-SFG, at different electrochemical potentials *versus* Pd/H$_2$.** (a) Experimental setup for the *in situ* electrochemical HD-SFG measurements. Monolayer graphene, hydrogen-loaded palladium wire, and gold wire are used as the WE, RE, and CE, respectively. (b) The O-H stretching $\text{Im}(\chi_{\text{eff}}^{(2)})$ spectra. We used 1 mM NaClO$_4$ aqueous solution. (c) The differential $\text{Im}(\Delta\chi_{\text{eff}}^{(2)})$ spectra obtained from taking the difference between $\chi_{\text{eff}}^{(2)}$ recorded at 1 mM and 100 mM NaClO$_4$ aqueous solutions. (d) Potential-dependent charge density of the graphene (from Raman spectra), the total surface charge at the CaF$_2$-supported graphene/water interface (from $\text{Im}(\Delta\chi_{\text{eff}}^{(2)})$) and inferred surface charge of the CaF$_2$ substrate (from the difference). (e) Raman spectra of the graphene on CaF$_2$ substrate, recorded under various applied potentials. The absence of the D band (~1350 cm$^{-1}$) demonstrates that the graphene remains electrochemically intact within the present electrochemical window. The spectra in b, c, and e are offset for clarity. The dashed lines in (b-d) indicate the zero line.

The flipping of the H-bonded O-H group orientations of the interfacial water molecules when we decrease the potential from -0.03 V to -0.63 V implies a drastic change in the surface charge of the CaF$_2$-supported graphene ($\sigma_{\text{net}}$). We can spectroscopically quantify the potential-dependent $\sigma_{\text{net}}$ from the $\chi_{\text{eff}}^{(2)}$ spectra using the electric double-layer model. The $\chi_{\text{eff}}^{(2)}$ signal at the charged interface can be modeled with Stern layer contribution ($\chi^{(2)}$-term) and the diffuse layer

contribution ($\chi^{(3)}$-term) via[25,37,38]

$$\chi_{\text{eff}}^{(2)}(\sigma_{\text{net}}, c) = \chi^{(2)} + \chi^{(3)}\phi_0(\sigma_{\text{net}}, c)\kappa(c)/(\kappa(c) - i\Delta k_z), \tag{1}$$

where $\chi^{(3)}$ represents the third-order nonlinear susceptibility originating from bulk water, $\phi_0$ is the electrostatic potential, $\kappa$ the inverse of Debye screening length, $c$ electrolyte concentration, and $\Delta k_z$ the phase-mismatch of the SF, visible, and IR beams in the depth direction. To obtain $\sigma_{\text{net}}$, we measured the $\chi_{\text{eff}}^{(2)}$ spectra at $c = 1$ mM and 100 mM, and then computed $\Delta\chi_{\text{eff}}^{(2)}(\sigma_{\text{net}}) = \chi_{\text{eff}}^{(2)}(\sigma_{\text{net}}, c = 1 \text{ mM}) - \chi_{\text{eff}}^{(2)}(\sigma_{\text{net}}, c = 100 \text{ mM})$. As such, one can omit the $\chi^{(2)}$-term and then connect $\Delta\chi_{\text{eff}}^{(2)}$ spectra with the total surface charge density, $\sigma_{\text{net}}$, through the $\chi^{(3)}$-term. (See *Supporting Information Section 6*).[24]

The $\text{Im}\left(\Delta\chi_{\text{eff}}^{(2)}\right)$ spectra are shown in Fig. 1c, from which we obtain $\sigma_{\text{net}}$ as a function of the applied potential (Fig. 1d). The $\sigma_{\text{net}}$ decreases gradually by 2 mC/m² when we change the potential from +0.57 V to -0.03 V. In the potential ranges from -0.03 V to -0.63 V, indeed, the $\sigma_{\text{net}}$ drops by 29 mC/m² via crossing the net neutral surface charge density at ~-0.33 V. Below -0.63 V, the $\sigma_{\text{net}}$ again varies relatively slowly against the applied potentials with the negative sign (-2 mC/m² changes from -0.63 V to -1.23 V).

Considering that supported graphene is known to be, at least in part, transparent in terms of substrate-water interactions,[39,40] the question arises whether this drastic variation of the surface charge arises from the graphene or the CaF$_2$ substrate. To address this question, we use Raman spectroscopy to independently determine the graphene Fermi level and, thereby its surface charge density ($\sigma_g$) (*see Supporting Information Section 7*).[41,42] With known $\sigma_{\text{net}}$ and $\sigma_g$, the surface charge density of the CaF$_2$ substrate ($\sigma_{\text{CaF}_2}$) can be obtained via

$$\sigma_{\text{CaF}_2} = \sigma_{\text{net}} - \sigma_g. \tag{2}$$

The Raman G band data is shown in Fig. 1e, while the inferred $\sigma_g$ is shown in Fig. 1d. The $\sigma_g$ varies nearly linearly against the potentials, consistent with references,[41,43] but in contrast with the variation of $\sigma_{\text{net}}$ obtained from the $\Delta\chi_{\text{eff}}^{(2)}$ spectra.

This result indicates that the drastic variation of the charge density of the CaF$_2$-supported graphene/water interface is caused by the change in the surface nature of CaF$_2$ substrate itself. Taking the $\sigma_g$ together with the $\sigma_{\text{net}}$, we deduced the $\sigma_{\text{CaF}_2}$ using Eq. (2). Fig. 1d shows that the $\sigma_{\text{CaF}_2}$ varies upon changing the potential applied to the graphene WE. At open circuit condition

(OCP, ~ +0.13V), the $\sigma_{CaF_2}$ is +21.5 mC/m². It changes very slightly when we change the potential between +0.57 V and -0.03 V, while it decreases drastically from -0.03 V to -0.63 V. Below -0.63 V, the $\sigma_{CaF_2}$ is nearly zero and is again insensitive to the applied potentials.

How does the charge density on the CaF$_2$ substrate change so dramatically by varying the potential between -0.03 V and -0.63 V? To obtain further insight into the interfacial structure at the graphene/water interface, we performed cyclic voltammetry (CV) measurements. The CV curve in Fig. 2a shows two typical features: the HER region (below -0.03 V, $H_2O \rightarrow H^+ + OH^-$)[44,45] and the double layer region (above -0.03 V). In the potential region of HER (*for details, see Supporting Information Section 8*), the generated OH$^-$ ions are expected to elevate the electrolyte solution's pH in the vicinity of graphene,[15,16] causing hydroxylation of the CaF$_2$ surface.

The CaF$_2$ substrate is positively charged at neutral pH (~5.6), with its isoelectric point in the range pH 9 to 10.[46,47] The adsorption of hydroxide ions can rationalize the variation of the charge on the CaF$_2$ substrate with pH, where ≡ indicates surface-bound:

$$\equiv Ca^+F \cdots H_2O \rightleftharpoons \,\equiv Ca(OH)F + H^+. \qquad (I)$$

The proton generated at the CaF$_2$/graphene interface can readily penetrate through the single-layer graphene to reach the bulk solution.[48–50]

$$H^+(CaF_2/graphene) \rightleftharpoons H^+(graphene/water). \qquad (II)$$

Remarkably, the potential of zero charge on the CaF$_2$ substrate is observed slightly above -0.33 V (Fig. 1d), implying that at this very moderate potential, already substantial hydroxide formation occurs, sufficient to increase the pH from 5-6 to 9-10. The pH change occurs at a remarkably low current density below 1 $\mu A/cm^2$ (Fig. 2a). To examine this hypothesis, we measured the $\text{Im}(\chi^{(2)}_{\text{eff}})$ spectra under various pH conditions at the CaF$_2$-supported graphene/water interface. The data is displayed in Fig. 2b. At bulk pH lower than 9.5, the spectra exhibit a negative H-bonded O-H band. When elevating the bulk pH above 9.5, the 2950-3550 cm$^{-1}$ band changes the sign from negative to positive, and a sharp negative peak at around 3630 cm$^{-1}$ appears, consistent with an isoelectric point between pH 9 and 10.

The spectral lineshapes of the O-H stretching mode at various pH conditions closely resemble those under applied potentials; the spectra at positive potentials and at low pH commonly show the negative 2950-3550 cm$^{-1}$ band, while comparing the spectra at negative potentials and high pH reveals the common features of a positive 2950-3550 cm$^{-1}$ band and a negative 3630 cm$^{-1}$ peak. In fact, the $\text{Im}(\chi_{\text{eff}}^{(2)})$ spectra at potentials of +0.17 V, -0.23 V, and -1.23 V perfectly overlap with the spectra at pH=5.6, 8.0, and 10.3, as seen in Figs 2c, 2d, and 2e, respectively. Clearly, the applied potential on the graphene causes the onset of HER, and the subsequent local pH change alters the surface charge density of $CaF_2$, which is responsible for the change in the water organization on the graphene surface.

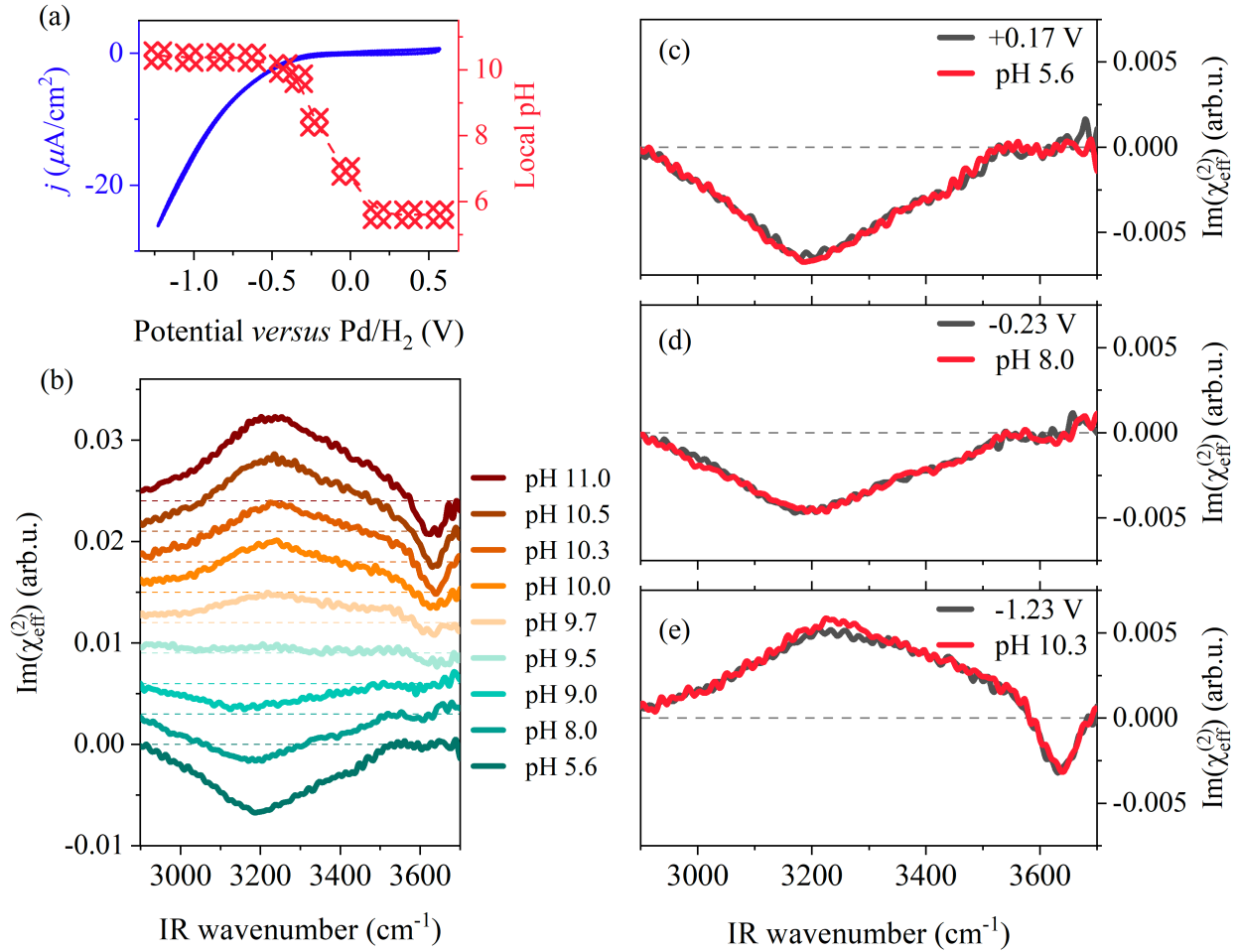

**Figure 2. Local pH change at the $CaF_2$-supported graphene/water interface.** (a) Cyclic voltammogram of the graphene electrode on $CaF_2$. We used 1 mM $NaClO_4$ aqueous solution, and the scan rate was set to 100 mV/s. Deduced local pH values at various applied potentials are also plotted. (b) The O-H stretching $\text{Im}(\chi_{\text{eff}}^{(2)})$ spectra near the graphene electrode supported on $CaF_2$ substrates at various pH conditions. These spectra are offset by 0.003 for clarity. (c, d, and e) Comparison of the applied potential effect and pH effect on the O-H stretching $\text{Im}(\chi_{\text{eff}}^{(2)})$ spectra.

The obtained correspondence of the applied potential and local pH is shown in Fig. 2a, which is correlated according to the integrated peak intensity of the H-bonded O-H peak and the Ca-O-H peak (*see the details in the Supporting Information Section 9*). The inferred pH value saturates at increasingly low potential. We expect the pH to increase continuously with increasing negative potential, but below -0.63 V, the SFG signal saturates. We tentatively assign the observed saturation to the limited number of water molecules confined between the graphene and $CaF_2$ as a reactant in chemical reaction (I), limiting that reaction. In any case, it is evident that applying a negative potential to the graphene induces the HER, elevating the local pH. The increase in local pH promotes the deprotonation of the water molecules confined in the $CaF_2$/graphene interface. This molecular picture is schematically depicted in Fig. 3a.

This mechanism can explain the sudden appearance of the Ca-O-H peak at around -0.43 V. At this potential, HER occurs, rapidly increasing the local pH above 10, promoting Ca-O-H formation through interfacial chemistry equilibria (I) and (II). This rapid increase in the local pH can be ascribed to the local condensation of $OH^-$ ions between $CaF_2$ and graphene because, unlike $H^+$,[51,53] $OH^-$ ions cannot penetrate through the graphene to the bulk water. The formation of Ca-O-H rapidly saturates at -0.63 V, further suggesting that the discharging of the $CaF_2$ substrate is limited by the amount of confined water between the graphene and the $CaF_2$ substrate. We note that this HER-induced local pH change and its dominant role in affecting the arrangement of interfacial water are also observed when the graphene electrode is supported on a silicon dioxide ($SiO_2$) substrate (see *Supporting Information Section 10*).

The local pH change rates at various applied potentials are different. We tracked the local pH change rate according to the peak area of the H-bonded O-H group (*see details in Supporting Information Section 9*). As shown in Fig. 3b, the local pH takes more than 900 seconds and 480 seconds to reach a steady-state condition at the potential of -0.23V and -0.33 V, but it takes only 120 seconds and less than 60 seconds at -0.43 V and -0.83 V, respectively. These varied local pH change rates at different potentials result from the different HER rates in the graphene/water interfacial region. The more negative the potential, the faster the HER and the faster chemical equilibria (I) and (II) are shifted to the right. The potential-dependent local pH change rates manifest that the local pH change is due to the occurrence of HER.

Finally, this process is reversible and reproducible (Fig. 3c). At positive potentials, proton transfer across the monolayer graphene causes fluoride ion dissociation, leaving a positively charged CaF$_2$ surface. Applying a negative potential drives HER to occur. Protons at the interface are rapidly consumed, and the fluoride ion dissociation is suppressed, leaving OH$^-$ ions at the CaF$_2$ surface, thereby discharging the CaF$_2$ surface. More importantly, by reversing the applied potential from negative to positive, the charging of the CaF$_2$ surface can again be triggered, although in a slower manner.

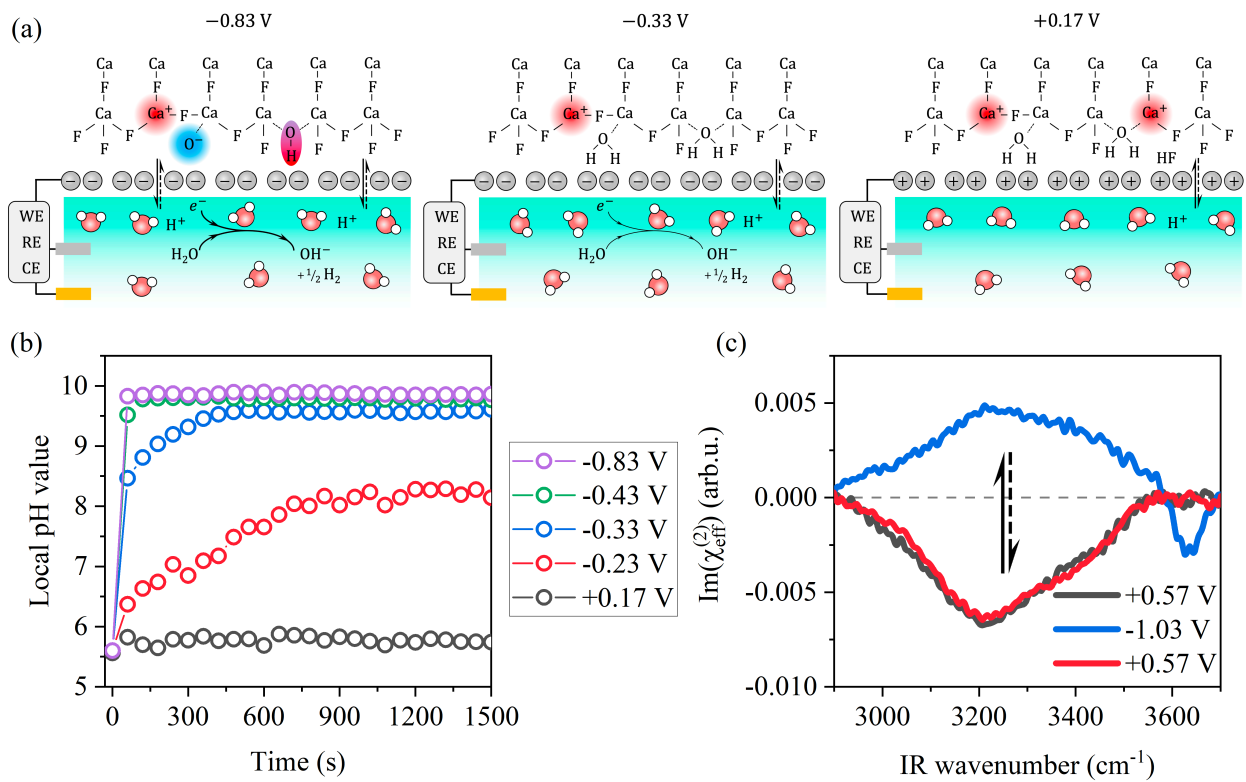

**Figure 3. Interfacial chemistry equilibria at the CaF$_2$-supported graphene/water interface due to local pH change.** (a) Schematic diagram of the CaF$_2$-supported graphene/water interface in a 1 mM NaClO$_4$ aqueous solution. At +0.17V, no chemical reaction occurs at the graphene/water interface, interfacial fluoride ions dissociation causes a positively charged CaF$_2$ surface. At -0.33 V, HER starts to occur, and OH$^-$ ions accumulate at the graphene/water interface, elevating the local pH. Interfacial fluoride ions dissociation was subsequently suppressed, and the positive surface charges on CaF$_2$ reduced. At more negative potentials, -0.83 V for example, interfacial chemical equilibria (I) and (II) shift to the right, causing a nearly neutral CaF$_2$ surface and the formation of Ca-O-H. (b) Local pH change rates at various applied potentials. The solid lines in the figure are to guide the eye. (c) Reversible surface charging/discharging of the CaF$_2$ substrate. Note that the Im$\left(\chi_{\text{eff}}^{(2)}\right)$ spectrum changes instantaneously when the potential changes from +0.57 V (black line) to -1.03 V (blue line), but it takes more than two hours from -1.03 V to 0.57 V (red line), probably due to the limited reaction rate of interfacial chemistry equilibria (I) and (II) to the left.

## Conclusion

We employed surface-specific spectroscopy to study water at the CaF$_2$-supported electrified graphene/water interface, and gain molecular-level insight into the chemical reaction and interfacial water molecules' arrangement near the electrified graphene. We found that the charges of the CaF$_2$ substrate change drastically between -0.63 V and -0.03 V, while they do not change in the regions of -1.23 V–-0.63 V and -0.03 V–+0.57 V. The variation of the SFG spectra under the applied potentials resembles the variation of the SFG spectra under varying pH conditions. This clearly manifests that the applied potential on the graphene triggers the HER, changing the local pH, which subsequently induces the discharging of the CaF$_2$ substrate. The deprotonation of the confined water between the graphene sheet and the CaF$_2$ substrate controls the charge of the CaF$_2$ substrate. The surface charge on the CaF$_2$ substrate varies with the applied potential, altering the interfacial electric field and, subsequently the arrangement of interfacial water at the graphene electrode surface. Finally, we note that this phenomenon is reversible and reproducible, and is general at the supported-graphene/water interface.

## Acknowledgment

We are grateful for the financial support from the MaxWater Initiative of the Max Planck Society. Y.K.W thanks China Scholarship Council for the support. Y.K.W also thanks the support from the Scientific Research Foundation of Graduate School of Southeast University (YBPY1946).

## Associated Content

More details about the sample preparation, electrochemical cell, Raman measurement, and HD-SFG measurement can be found in the *Supporting Information 1-4*, respectively. Wettability of monolayer graphene on CaF$_2$, a procedure to extract the $\sigma_{\text{net}}$ from $\chi_{\text{eff}}^{(2)}$ spectra, a procedure to extract the $\sigma_{\text{g}}$ from Raman spectra, discussion on the hydrogen evolution reaction, a procedure to calculate the local pH, and the $\text{Im}\left(\chi_{\text{eff}}^{(2)}\right)$ spectra measured at the SiO$_2$-supported graphene/water interface are also given in *Supporting Information 5-10*, respectively.

# Supporting Information

# Surface-specific vibrational spectroscopy of interfacial water reveals large pH change near graphene electrode at low current densities

*Yongkang Wang,[1,2,#] Takakazu Seki,[2,#] Xuan Liu,[3] Xiaoqing Yu,[2] Chun-Chieh Yu,[2] Katrin F. Domke,[2,4] Johannes Hunger,[2] Marc T. M. Koper,[3] Yunfei Chen,[1,*] Yuki Nagata,[2,*] and Mischa Bonn[2,*]*

[1] School of Mechanical Engineering, Southeast University, 211189 Nanjing, China.
[2] Max Planck Institute for Polymer Research, Ackermannweg 10, 55128 Mainz, Germany.
[3] Leiden Institute of Chemistry, Leiden University, Einsteinweg 55, 2333CC Leiden, The Netherlands.
[4] University Duisburg-Essen, Faculty of Chemistry, Universitätsstraße 5, 45141 Essen, Germany.
[#] Yongkang Wang and Takakazu Seki contributed equally to this work.

*Correspondence to: yunfeichen@seu.edu.cn, nagata@mpip-mainz.mpg.de, bonn@mpip-mainz.mpg.de

## Supporting Information 1    Sample preparation

The sodium hydroxide (NaOH), hydrochloride (HCl), ammonium persulfate (($NH_4$)$_2$$S_2$$O_8$), concentrated sulfuric acid ($H_2SO_4$), 30 wt. % hydrogen peroxide solution ($H_2O_2$), cellulose acetate butyrate (CAB), ethyl acetate, ethanol, and acetone (> 99%, analytical grade) were purchased from Sigma-Aldrich, and used without further purification. Sodium perchlorate ($NaClO_4$, metals basis, 99.99%) was obtained from Merck, and used as a supporting electrolyte in our electrochemical experiment with a concentration of 1 mM, 10 mM or 100 mM. Deionized water (pH ~5.6) was provided by a Milli-Q system (resistivity ≥ 18.2 MΩ·cm and TOC ≤ 4 ppb). A CVD-grown monolayer graphene on a copper foil was purchased from Grolltex Inc.

Prior to transferring the graphene, calcium fluoride ($CaF_2$) and fused silica ($SiO_2$) windows (25 mm diameter with a thickness of 2 mm, PI-KEM Ltd.) were ultrasonically cleaned in acetone, ethanol, and deionized water sequentially for five minutes in each process. Subsequently, the $CaF_2$ windows were baked at 450 °C for 5 hours, and then immersed in an HCl solution at pH 2 for 2 hours to generate the fresh $CaF_2$ surface. Different from the $CaF_2$ windows, the $SiO_2$ windows were immersed in a piranha solution for 15 minutes before use. After that, two 100 nm-thick gold strips were thermally evaporated onto the $CaF_2$ and $SiO_2$ windows with a shadow mask. The gold strips enable us to measure the resistance across the graphene electrode (typical resistance of the graphene monolayer is 0.5–1.5 kΩ at a size of 6×12 $mm^2$) and to manipulate electrochemical potentials between the graphene and the reference electrode. Besides, the gold strips also serve as the reference sample to generate a stable and precise reference phase.

The CVD-grown monolayer graphene was transferred onto the cleaned $CaF_2$ and $SiO_2$ windows by using the polymer-assisted wet transfer technique.[1] In brief, the copper foil was spin-coated with CAB (30 mg/mL dissolved in ethyl acetate) at 1,000 rpm for 10 seconds, followed by 4,000 rpm for 60 seconds and then baked at 180 °C for 3 minutes. After cooling down to room temperature, the copper foil/CAB was placed into $HCl/H_2O_2/H_2O$ mixture solution (volume ratio, 1:1:10) for 60 seconds to remove the graphene layer grown on the backside of the copper foil. After being rinsed with deionized water, the copper foil/CAB was then etched away in 0.1 M ammonium persulfate aqueous solution. Subsequently, the obtained CAB-graphene films were rinsed in deionized water several times to remove residual chemical species, and then were transferred onto $CaF_2$ and $SiO_2$ windows. The samples were dried for more than 12 hours at 110 °C in a vacuum (~1 mbar) to remove residual water. Finally, the CAB layer on graphene was dissolved in acetone. To ensure good electrical contact, two gold wires were bound to the two gold strips on the $CaF_2$ and $SiO_2$ window with conductive silver paste (TED PELLA, INC.).

*Supporting Information 2    Electrochemical cell*

Our electrochemical flowing liquid cell is schematically depicted in Fig. 1a. The cell mainly consists of two rectangular polytetrafluoroethylene (PTFE) parts, the top clamp part, and the bottom flowing channel (12×3×3 $mm^3$) part. The top clamp has an opening of ~16 mm in diameter for the light beam paths. The bottom part has four round holes on the four side walls. Two for inlets and outlets of electrolyte solution (1 mM, 10mM or 100 mM $NaClO_4$) while another two for insertion of the hydrogen-loaded palladium wire (RE) and the gold wire (CE). The monolayer graphene electrode on a $CaF_2$ or a $SiO_2$ window and an O-ring were then sandwiched between the top and the bottom PTFE parts. The O-ring was used to create a seal between the electrolyte solution and the graphene electrode to avoid contact between the the solution and the two gold strips. The base and clamp parts were cleaned with piranha solution before use.

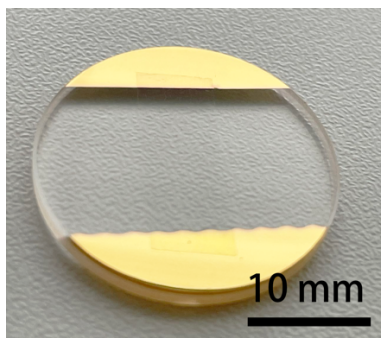

**Figure S1.** Optical image of the transferred monolayer graphene on a CaF$_2$ substrate. Two gold strips were thermally deposited onto the CaF$_2$ window before the graphene transfer for electrical connection.

In our electrochemical experiment, a gold wire (0.5 mm in diameter, Alfa Aesar, Premion, 99.9985% metals basis) served as the CE, and was freshly flame-annealed and rinsed with water before use. The Pd wire (0.5 mm in diameter, MaTecK, 99.95% metals basis), after being freshly flame-annealed, was loaded with hydrogen via immersion into 0.1 M H$_2$SO$_4$ with an applied voltage of 5 V between the Pd wire and a fresh-prepared gold wire for about 10 minutes. The monolayer graphene, gold wire, and hydrogen-loaded palladium wire (Pd/H$_2$) electrodes served as the WE, CE, and RE, respectively. These electrodes were connected to an electrochemical workstation (Metrohm Autolab PGSTAT302).

*Supporting Information 3    Raman measurement*

The Raman spectra of graphene samples under various electrochemical potentials were recorded with a WITec confocal Raman spectrometer (alpha 300 R, ×10 objective, 600 grooves/mm grating, 5 mW) with a 532 nm excitation and 10 seconds integration. The G-band peak frequency was obtained by fitting it with Lorentzian curves after the background subtraction.

*Supporting Information 4    SFG measurement*

HD-SFG measurements were performed on a non-collinear beam geometry with a Ti:Sapphire regenerative amplifier laser system (Spitfire Ace, Spectra-Physics, centered at 800 nm, ~40 fs pulse duration, 5 mJ pulse energy, 1 kHz repetition rate). A part of the output was directed to a grating-cylindrical lens pulse shaper to produce a narrowband visible pulse (10 µJ pulse energy, FWHM = ~10 cm$^{-1}$), while the other part was used to generate a broadband infrared (IR) pulse

(3.5 μJ pulse energy, FWHM = ~530 cm$^{-1}$) through an optical parametric amplifier (Light Conversion TOPAS-C) with a silver gallium disulfide (AgGaS$_2$) crystal. The IR and visible beams were firstly focused into a 200 nm-thick ZnO on a 1 mm-thick CaF$_2$ window to generate a local oscillator (LO) signal in a similar manner to the reference.[2] Then, IR, visible, and LO beams were re-focused by two off-axis parabolic mirrors pair and overlapped spatially and temporally at the graphene/water interface at the angles of incidence of 33°, 39°, and 37.6°, respectively. A fused silica glass plate with a 1.5 mm thickness was placed in the optical path for the LO signal in between the two off-axis parabolic mirrors pair, allowing the phase modulation for the LO signal. The SFG signal from the sample interfered with the SFG signal from the LO, generating the SFG interferogram, which was then dispersed in a spectrometer (Shamrock 303i, Andor Technology) and detected by an EMCCD camera (Newton, Andor Technology). HD-SFG spectra were measured in an N$_2$ atmosphere to avoid spectral distortion due to water vapor. The measurement was conducted at *ssp* polarization combination, where *ssp* denotes *s*-polarized SFG, *s*-polarized visible, and *p*-polarized IR beams. To avoid erroneous measurements due to the change of the surface height of the sample upon flowing electrolyte solutions, we used a height displacement sensor (CL-3000, Keyence). Each spectrum was acquired with an exposure time of 15 minutes, and measured three times for average.

The complex-valued spectra of second-order nonlinear susceptibility ($\chi_{\text{eff}}^{(2)}$) of the graphene/water interface samples were obtained via the Fourier analysis of the interferogram and normalization with that of the CaF$_2$/gold interface. The interferogram of the CaF$_2$/gold interface was collected at the gold strips region of the sample immediately before the sample measurement to ensure a precise and stable reference phase. The procedures are described below. The total intensity of the signal light measured in the HD-SFG is represented by the sum of sample sum frequency (SF) light and local oscillator (LO) reflected at an interface, which can be expressed as follows:

$$|E_{\text{total}}|^2 = |E_{\text{sample}}|^2 + |E_{\text{LO}}|^2 + E_{\text{sample}}E_{\text{LO}}^* e^{i\omega T} + E_{\text{sample}}^* E_{\text{LO}} e^{-i\omega T}, \quad (S1)$$

where $E_{\text{sample}}$ and $E_{\text{LO}}$ are the electric fields of the SF light from the sample and the LO, respectively, with * representing the complex conjugate. $T$ is the time delay between SF lights from the sample SF and from the LO. The third term in the right-hand side of Eq. S1 was picked up using time-domain filtration with Fourier transform to obtain $E_{\text{sample}}$ contribution. For *ssp* polarization combination, the $\chi^{(2)}_{\text{eff}}$ spectrum of the graphene/water interface in the O-H stretching region is obtained via

$$\chi^{(2)}_{\text{eff}} = \frac{E_{\text{water}} E^*_{\text{LO}} \exp(i\omega T)}{E_{\text{gold}} E'^*_{\text{LO}} \exp(i\omega T)}, \qquad (S2)$$

For the comparison of the spectra in this work, we assume the reflectivity for the electrolyte solution/graphene interfaces is insensitive to the type of electrolyte and concentration below 100 mM.

Note that the absolute phase of the gold thin film is not uniquely determined. To obtain the phase of the gold sample, we measured the O-H stretching $\text{Im}(\chi^{(2)}_{\text{eff}})$ spectrum of the $CaF_2$-supported graphene/$D_2O$ interface via the normalization with HD-SFG interferogram of $CaF_2$/gold. As $D_2O$ does not have any vibrational response in this region, and its $\chi^{(2)}_{\text{eff}}$ response arises solely from the interface[3], we can determine the phase of gold based on the fact that the $\text{Im}(\chi^{(2)}_{\text{eff}})$ spectrum of the $CaF_2$-supported graphene/$D_2O$ interface shows a flat zero line (Fig. S2).

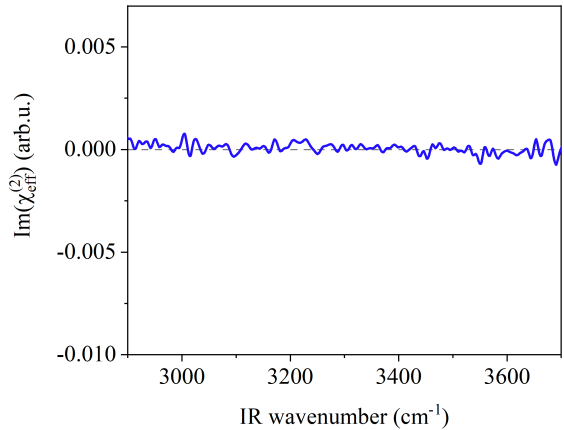

**Figure S2. The O-H stretching $\text{Im}(\chi^{(2)}_{\text{eff}})$ spectra obtained from the $CaF_2$-supported graphene/$D_2O$ interfaces at -1.03 V.** The dashed line indicates the zero line.

## *Supporting Information 5    Wettability of monolayer graphene on $CaF_2$*

The $\text{Im}(\chi^{(2)}_{\text{eff}})$ spectra of the $CaF_2$/water interface in the presence and absence of graphene are shown in Fig. S3. The data clearly shows that the presence of the monolayer graphene weakly affects the substrate-water interaction, consistent with the fact that graphene is known as wetting transparent.[4,5]

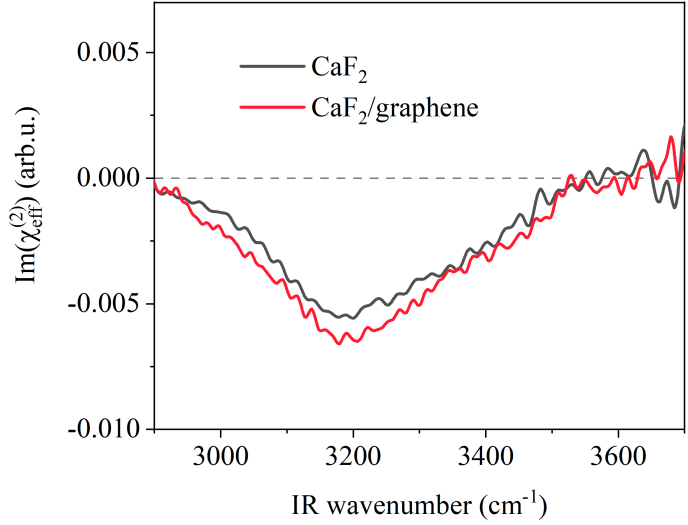

**Figure S3. Transparency of monolayer graphene in terms of substrate-water interaction.** The O-H Im$\left(\chi_{\text{eff}}^{(2)}\right)$ spectra at the CaF$_2$/water interfaces in the presence and absence of the monolayer graphene. The spectra were measured at OCP condition. The dashed line serves as the zero line.

## *Supporting Information 6  Surface charge density at the CaF$_2$-supported graphene/water interface, $\sigma_{net}$*

We extract the $\sigma_{\text{net}}(V)$ at applied potential $V$ from the $\chi_{\text{eff}}^{(2)}$ spectra following the procedures developed by Shen et al. and Bonn et al.[6,7] In brief, we measured the $\chi_{\text{eff}}^{(2)}$ spectra at $c$ = 1 mM (Fig. 1b) and 100 mM (Fig. S4), and then computed the differential spectra $\Delta\chi_{\text{eff}}^{(2)}(\sigma_{\text{net}}, V) = \chi_{\text{eff}}^{(2)}(\sigma_{\text{net}}(V), c_1 = 1 \text{ mM}) - \chi_{\text{eff}}^{(2)}(\sigma_{\text{net}}(V), c_2 = 100 \text{ mM})$. The $\text{Im}\left(\Delta\chi_{\text{eff}}^{(2)}\right)$ at various potentials are shown in Fig. 1c. Under the assumption that the effect of the ions on the $\chi^{(2)}$ at low enough concentration (1–100 mM) is negligible, the amplitude of $\Delta\chi_{\text{eff}}^{(2)}$ is mainly related to the difference of $\chi^{(3)}$ terms at different concentrations. By dividing $\Delta\chi_{\text{eff}}^{(2)}$ with $\chi^{(3)}$, we further obtained

$$\frac{\Delta\chi_{\text{eff}}^{(2)}(\sigma_{\text{net}}(V), c_1, c_2)}{\chi^{(3)}} = \frac{\phi_0(\sigma_{\text{net}}, c_1)\kappa(c_1)}{\kappa(c_1) - i\Delta k_z} - \frac{\phi_0(\sigma_{\text{net}}, c_2)\kappa(c_2)}{\kappa(c_2) - i\Delta k_z}. \tag{S3}$$

We compared the experimentally obtained left side of Eq. S3 and computed right side, and subsequently obtained the $\sigma_{\text{net}}$. Here, to compute the left side, we used $\chi^{(3)}$ spectrum obtained from the reference[7] to improve signal-to-noise ratio. Note that the spectral shape between the experimentally obtained $\text{Im}(\chi^{(3)})$ spectrum in this work and that reported in Ref. [7] perfectly matched (Fig. S5). Thus, this operation does not affect the results of the surface charge determinations. The obtained $\sigma_{\text{net}}$ at various potentials are displayed in Fig. 1d.

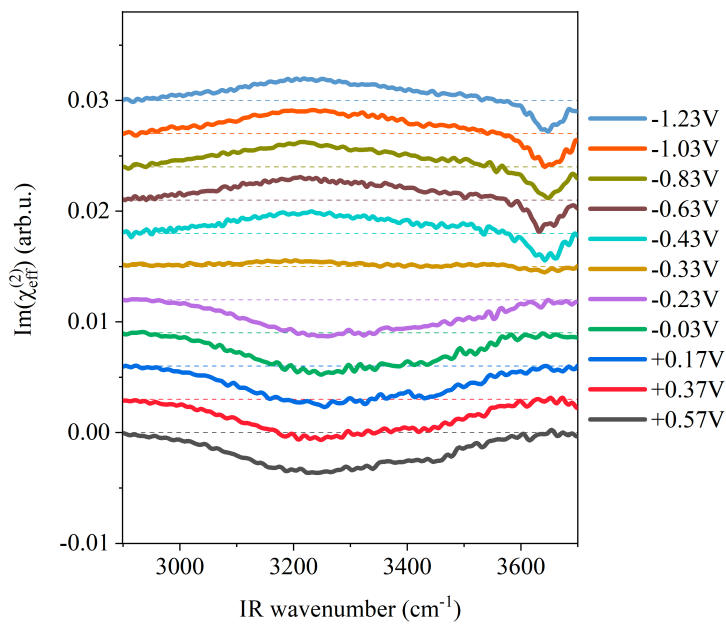

**Figure S4. O-H stretching $\text{Im}\left(\chi_{\text{eff}}^{(2)}\right)$ spectra at different potentials *vs.* Pd/H$_2$.** We used 100 mM NaClO$_4$ aqueous solution. The data is offset by 0.003 for clarity. The dashed lines indicate the zero line.

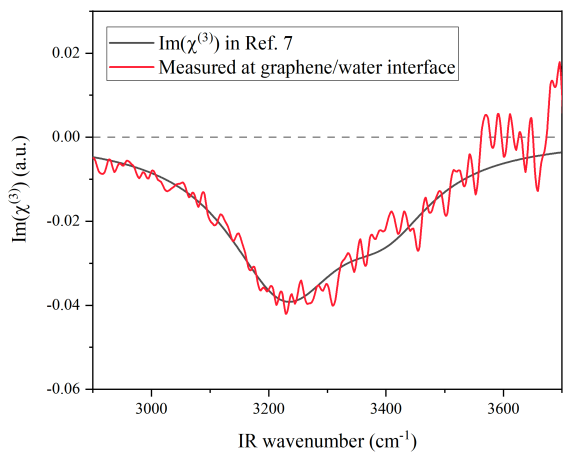

**Figure S5. Comparison of the measured $\text{Im}(\chi^{(3)})$ spectrum and that reported in Ref. 7.** The dashed line indicates the zero line.

Figure S6 shows the results we got at the OCP condition, which gives an estimation of 21.5 mC/m$^2$ net surface charge density. Since the charge density in graphene is nearly zero at the OCP condition, the 21.5 mC/m² surface charge mainly comes from the CaF$_2$ substrate.

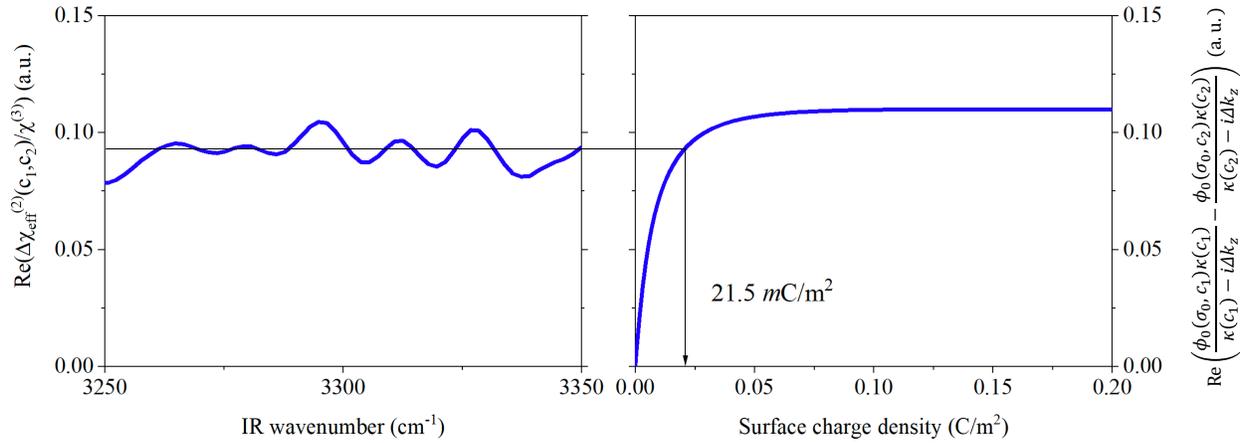

**Figure S6.** Net surface charges of the CaF$_2$/graphene/water interface at OCP condition obtained from $\Delta\chi_{eff}^{(2)}$ spectral analysis.

*Supporting Information 7    Charge density in graphene $\sigma_g$*

The application of a gate voltage injects charge carriers in graphene, leading to a shift in the Fermi level. The Fermi level in graphene ($E_F$) changes with the density of the charge carriers via $E_F = \hbar|v_F|\sqrt{\pi n}$, where $v_F (= 1.1 \times 10^{-6}$ ms$^{-1}$) is the Fermi velocity in monolayer graphene,[8] and $n$ denotes the charge carrier density in graphene. Meanwhile, it has been shown that the shift in the Raman G-band frequency linearly correlated with the Fermi level via $E_F = 21\Delta\omega_G + 75$ [cm$^{-1}$] for electrons and $E_F = -18\Delta\omega_G - 83$ [cm$^{-1}$] for holes carriers.[9,10] The charge density in graphene ($\sigma_g$) at various potentials can therefore be determined via Raman spectral analysis of the Raman G-band frequency. The obtained Raman spectra and the deduced surface charge densities are given in Fig. 1d. As we obtained the surface charge of the graphene in this section and that for the CaF$_2$/graphene water interface in the previous section, we can now deduce the surface charges on CaF$_2$ substrate ($\sigma_{CaF_2}$) via $\sigma_{CaF_2} = \sigma_{net} - \sigma_g$. The deduced $\sigma_{CaF_2}$ is showcased in Fig. 1d.

*Supporting Information 8    Hydrogen evolution reaction*

As discussed in the main text, the CV curve in Fig. 2a shows an electrochemical reduction peak starting below ~-0.23 V. In this potential region, two reduction reactions may occur, HER

($H_2O \rightarrow H^+ + OH^-$)[11,12] and ORR ($O_2 + 2H_2O + 4e^- \rightarrow 4OH^-$),[13] both of which can change the local pH. To understand which one dominates at the graphene electrode, we measured CV curve with argon-saturated solution. The data is shown in Fig. S7a. The two CV curves are the same, which suggests that the electrochemical reduction peak is due to HER.

Further, we measured CV curves under various pH conditions. We use $SiO_2$-supported graphene rather than $CaF_2$-supported graphene as the working electrode to avoid the detaching of graphene from the $CaF_2$ surface under acid conditions. The data is shown in Fig. S7b. Under acidic conditions (pH=2), the CV curve shows first a significant proton reduction peak at around -1.1V and then a water reduction peak at lower than -1.5 V. The proton reduction peak becomes much weak under neutral conditions due to the limited number of protons in the solution, and the peak nearly vanishes under alkaline conditions (pH=12). These results illustrate that applying negative potentials on graphene triggers the HER,[14] and the HER is responsible for the local pH change.

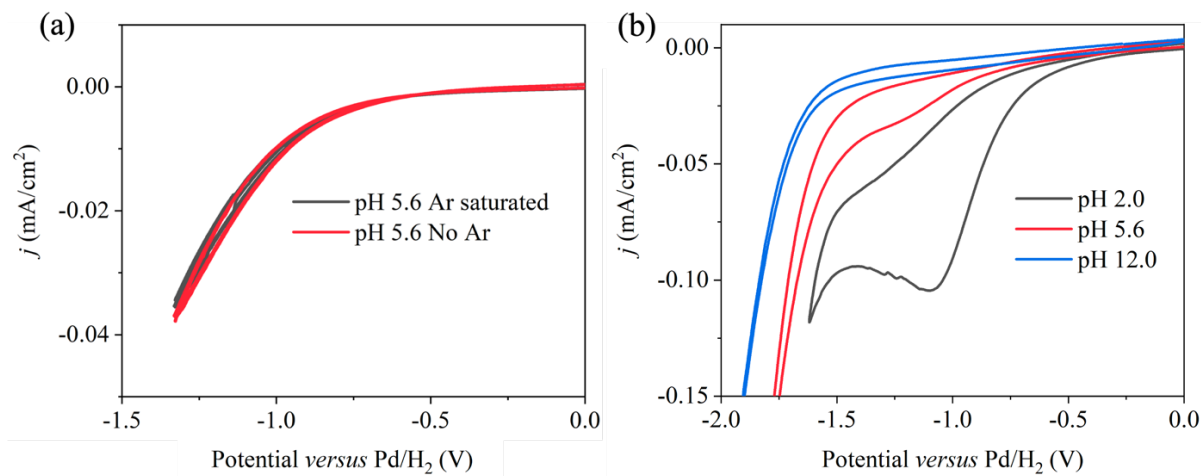

**Figure S7. Cyclic voltammogram of the graphene electrode.** (**a**) CV curves of graphene in contact with 1 mM $NaClO_4$ solution before and after argon-saturated treatment. The graphene electrode is supported on $CaF_2$ substrate. (**b**) CV curves of graphene in contact with 10 mM $NaClO_4$ solution at various pH conditions. The graphene electrode is supported on an $SiO_2$ substrate. We set the scan rate to 50 mV/s.

*Supporting Information 9     Determination of local pH value at various applied potentials*

As discussed in the main text, the lineshapes of the spectra of the O-H stretching mode at various pH conditions resemble those obtained upon applying potentials. As both the H-bonded O-H peak and Ca-O-H peak in the $\text{Im}(\chi^{(2)}_{\text{eff}})$ spectra will change when the surface charge density on the $CaF_2$ substrate is altered by pH, one can determine the local pH value by comparing the variations of the peak area of either the H-bonded O-H peak or the Ca-O-H peak by pH change and by potential change.

Here, to correlate them, first, we use the Henderson-Hasselbalch equation[15] to describe the relationship between the pH and the peak area of the H-bonded O-H peak as follows under the assumption that the peak area is proportional to the surface density of the charged species on CaF$_2$ surface:

$$\text{Area}(pH) = \frac{\text{Area}_{max} + \text{Area}_{min} \times 10^{(pKa-pH)}}{1 + 10^{(pKa-pH)}} \quad (S4)$$

where $\text{Area}_{min} = -2.0$ and $\text{Area}_{max} = 5.2$ are the minimum and maximum values of the peak area, respectively. The fit yields a pKa of 9.6 (Fig. S8a). From this, one can calculate the corresponding local pH value at each potential once the peak area of the H-bonded O-H peak upon applying potential is known.

For potentials at -0.33 V– -1.23 V, the peak area of the of Ca-O-H peak at 3630 cm$^{-1}$ also provides a measurement of local pH. In the same way, we use Eq. S4 to describe the relationship between the pH and the peak area of the Ca-O-H peak. A reasonable fitting yield $\text{Area}_{min} = 0.0$, $\text{Area}_{max} = -0.3$ and pKa = 9.9 (Fig. S8b). one can also calculate the corresponding local pH value with the known peak area of the Ca-O-H peak.

The calculated local pH value at each applied potential using the above two methods gives the same results as shown in Fig. S9c. For potentials at -0.33 V– -1.23 V, we use the Ca-O-H peak to calculate the local pH value, while at -0.23V– +0.57V, we use the H-bonded O-H peak. The results are shown in Fig. 2a. When measuring the local pH change rates at various potentials (Fig. 3b), we always use the H-bonded O-H peak to calculate the local pH value.

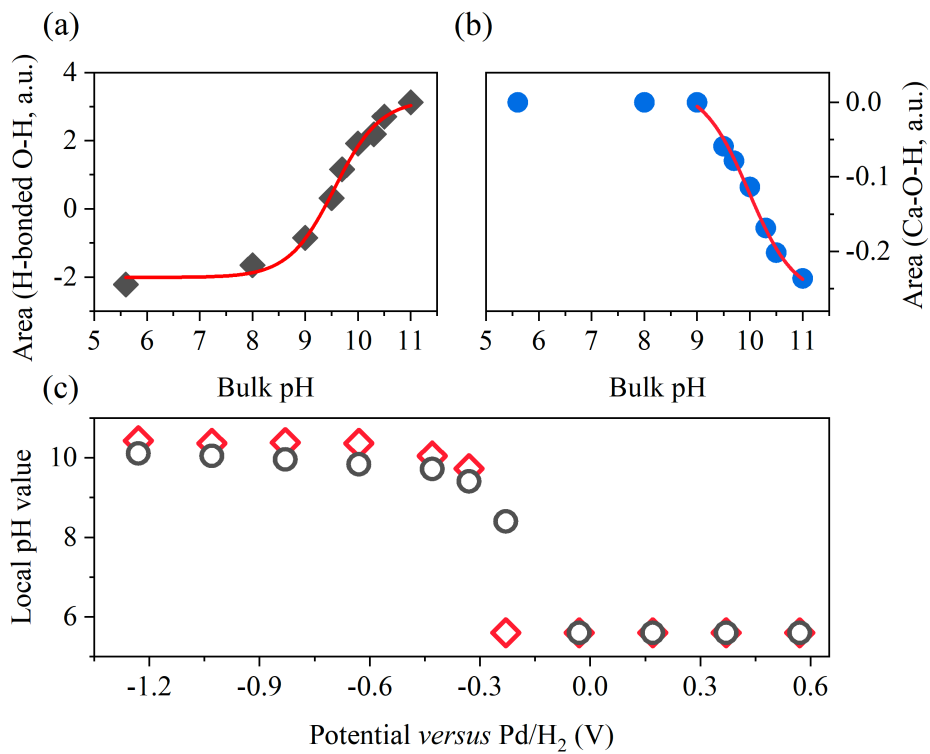

**Figure S8. Determination of local pH value at various potentials.** (**a**) Peak area of $\text{Im}(\chi^{(2)}_{\text{eff}})$ spectra in the H-bonded O-H region. The peak area of the Ca-O-H stretch peak at various pH is shown in (**b**). The red lines are fitting results using the Henderson-Hasselbalch equation. (**c**) Calculated local pH value at various potentials using the two methods.

## *Supporting Information 10   SiO$_2$-supported graphene electrode*

To further verify the local pH change near the graphene electrode, we measured the $\text{Im}(\chi^{(2)}_{\text{eff}})$ spectra at the SiO$_2$-supported graphene/water interface. The data is shown in Fig. S10. Different from the CaF$_2$-supported graphene, the $\text{Im}(\chi^{(2)}_{\text{eff}})$ spectra measured at the SiO$_2$-supported graphene/water interface exhibit a positive H-bonded O-H group at both positive and negative potentials. This is because the isoelectric point of SiO$_2$ is in the range of pH=2 to 3.[16] At +0.57 V (pH 5.6), the SiO$_2$ is negatively charged. Changing the voltage to -1.23V, HER-induced local pH increase (pH >10) will increase the negative surface charge density on SiO$_2$. Therefore, the positive H-bonded O-H group increase at -1.23V. More importantly, the $\text{Im}(\chi^{(2)}_{\text{eff}})$ spectrum measured at -1.23V overlaps with the spectrum measured at pH=10.5. These results demonstrate that the rearrangement of the interfacial water is due to the local pH change near the graphene electrode.

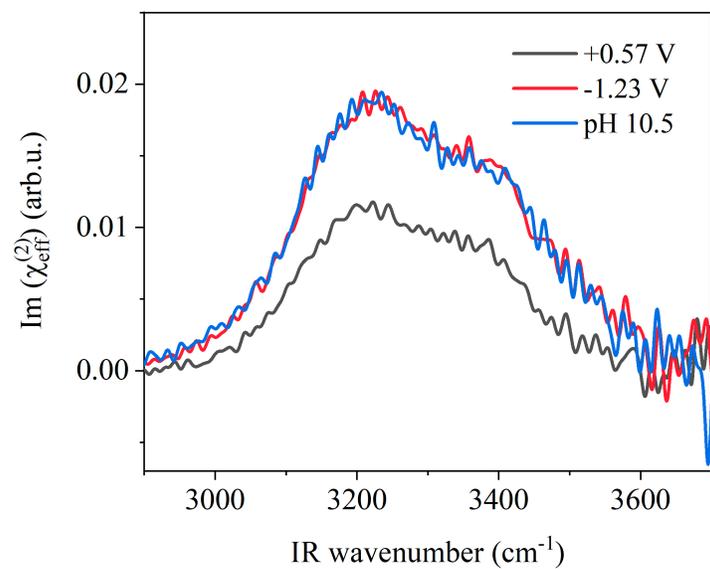

**Figure S10. O-H stretching Im $\left(\chi_{\text{eff}}^{(2)}\right)$ spectra at the SiO₂-supported graphene/water interface.** We used 1 mM NaClO₄ aqueous solution. The dashed line serves as the zero line.